\documentclass{elsart1p}

\usepackage{amssymb}
\usepackage{graphicx}

\newcommand{\beq}{\begin{equation}}
\newcommand{\eeq}{\end{equation}}
\newcommand{\bqa}{\begin{eqnarray}}
\newcommand{\eqa}{\end{eqnarray}}

\def\square{\vcenter{\vbox{\hrule height.4pt
          \hbox{\vrule width.4pt height4pt
          \kern4pt\vrule width.3pt}\hrule height.4pt}}}

\begin{document}

\begin{frontmatter}



\title{Bose-Einstein condensation in dense quark matter}


\author{Jens O. Andersen}

\address{Department of Physics, Norwegian University of Science and Technology,
N-7491 Trondheim, Norway}

\begin{abstract}
We consider the problem of Bose condensation of charged pions in QCD
at finite isospin chemical potential $\mu_I$
using the $O(4)$-symmetric linear sigma model as an effective field theory
for two-flavor QCD.
Using the 2PI $1/N$-expansion, we determine the quasiparticle masses
as well as the pion and chiral condensates as a function of the temperature
and isospin chemical potential in the chiral limit and at the physical point.
The calculations show that there is a competition between the condensates.
At $T=0$, Bose condensation takes place for chemical potentials
larger than $m_{\pi}$. In the chiral limit, the chiral condensate
vanishes for any finite value of $\mu_I$.

\end{abstract}

\end{frontmatter}

\section{Introduction}
Condensation of bosons occurs in various contexts of high-density
QCD. In the
color-flavor-locked (CFL) phase of high-baryon density 
QCD, the $SU(3)_c\times SU(3)_L\times SU(3)_R\times U(1)_B$
is broken down to $SU(3)_{c+L+R}$ and the breaking of the global symmetries
gives rise to (pesudo)-Goldstone bosons. The
lightest of the excitations are expected to be
the neutral and charged kaons. If
the chemical potentials are large enough and the temperature low enough, the
kaons will condense. Kaon condensation in the CFL phase
is discussed e.g. in Ref.~\cite{alfordus} and also in 
the talk given by Andreas Schmitt at this 
conference~\cite{andreas}.

Another interesting system in this context is QCD at finite isospin
chemical potential $\mu_I$. 
If the isospin chemical potential is large enough and the
temperature sufficiently low, the charged pions will condense.
In contrast to QCD at finite baryon chemical potential, QCD at finite
isospin chemical potential
is accessible on the lattice since there is no fermion sign problem. 
Lattice simulations~\cite{lattqcd}
indicate that there is a deconfinement transition of pions at high temperature
and low density, and Bose-Einstein condensation of charged pions at high
isospin density and low temperature. 
Various aspects of 
pion condensation have been investigated using chiral perturbation 
theory~\cite{cpt}, (Polyakov loop)NJL models~\cite{njlmod,china,pnjl}, and 
linear
sigma models~\cite{china,jens,jenstom}.

In this talk, I would like to present some results for Bose condensation 
of charged pions using the linear sigma model as a low-energy effective
for two-flavor QCD. The talk is based on Refs.~\cite{jens,jenstom} 
and part of the
work is done in collaboration with Tomas Brauner.

\section{Interacting Bose gas and 2PI $1/N$-expansion}
The Euclidean
Lagrangian for a Bose gas with $N$ species of massive charged scalars is
\bqa
{\cal L}&=&
(\partial_{\mu}\Phi_i^{\dagger})(\partial_{\mu}\Phi_i)
-{H\over\sqrt{2}}\left(\Phi_1+\Phi_1^{\dagger}\right)
+m^2\Phi^{\dagger}_i\Phi_i
+{\lambda\over2N}\left(\Phi^{\dagger}_i\Phi_i\right)^2
\;,
\label{lag}
\eqa
where $i=1,2,...,N$ and 
$\Phi_i=(\phi_{2i-1}+i\phi_{2i})/\sqrt{2}$ is a complex field.
A nonzero value of $H$ allows for explicit symmetry breaking and 
finite pion masses in the vacuum. 
The incorporation of a conserved charge $Q_i$ is done by making the 
substitution 
$\partial_0\Phi_i\rightarrow\left(\partial_0-\mu_i\right)\Phi_i$ in the 
Lagrangian~(\ref{lag}).
We introduce nonzero expectation values $\phi_0$ and $\rho$ 
for $\phi_1$ and $\phi_2$, respectively, to allow for a chiral
condensate and a charged pion condensate.
We then write the complex field $\Phi_1$ as
$\Phi_1={1\over\sqrt{2}}(\phi_0+i\rho_0+\phi_1+i\phi_2),$
where $\phi_1$ and $\phi_2$ are quantum fields.
The inverse tree-level propagator $D_0^{-1}$ reads
\bqa
D_0^{-1}
=
\left(\begin{array}{cccc}
\omega_n^2+p^2+m_1^2&{\lambda\over N}\phi_0\rho_0&0&0\vspace{2mm}
\\
{\lambda\over N}\phi_0\rho_0&\omega_n^2+p^2+m_2^2&-2\mu_I\omega_n&0
\\
0&2\mu_I\omega_n&\omega_n^2+p^2+m_3^2&0
\\
0&0&0&\omega_n^2+p^2+m_4^2 \\
\end{array}\right)\;,
\label{d0}
\eqa
where $p=|{\bf p}|$, $\omega_n=2n\pi T$ is the nth Matsubara frequency, and 
the treel-level masses are
\bqa
m_1^2=m^2+{3\lambda\over2N}\phi_0^2+{\lambda\over2N}\rho_0^2\;,\hspace{0.3cm}
m_2^2=-\mu^2_I+m^2+{\lambda\over2N}\phi_0^2+{3\lambda\over2N}\rho_0^2\;,\\
m_3^2=-\mu^2_I+m^2+{\lambda\over2N}\phi_0^2+{\lambda\over2N}\rho_0^2\;,
\hspace{0.3cm}
m_4^2=m^2+{\lambda\over2N}\phi_0^2+{\lambda\over2N}\rho_0^2\;.
\eqa
We have introduced a chemical potential $\mu_I$ for the third component of
the isospin charge, which corresponds to the current density
$j^{\mu}\sim\phi_2\partial^{\mu}\phi_3-\phi_3\partial^{\mu}\phi_2$.
For nonzero $\mu_I$, the charged pions $\pi^{\pm}$ are linear combinations of
$\phi_2$ and $\phi_3$.

After symmetry breaking, the 2PI effective potential can be written as
\bqa\nonumber
\Gamma[\phi_0,\rho_0,D]&=&
{1\over2}m^2\left(\phi_0^2+\rho_0^2\right)
+{\lambda\over8N}\left(\phi_0^2+\rho_0^2\right)^2
-{1\over2}\mu_I^2\rho_0^2
-H\phi_0
+{1\over2}{\rm Tr}\ln D^{-1}
\\ &&
+{1\over2}{\rm Tr}D_0^{-1}D
+\Phi[D]\;,
\label{ea1}
\eqa
where $D$ is the exact propagator and
$\Phi[D]$ is the sum of all two-particle irreducible vacuum diagrams.
The contributions to $\Phi[D]$ can be classified according to which order
in $1/N$ they contribute. The leading-order contribution is the double-bubble
diagram where the $2N-3$ particles with tree-level masses $m_4$
are propagating in the loops. 

The self-energy in the large-$N$ limit, $\Pi_{\rm LO}$, is given by the
tadpole diagram and 
is independent of momentum. The inverse propagator can then be written as
$D^{-1}=p^2+\omega_n^2+m_4^2+\Pi_{\rm LO}$. The term
$m_4^2+\Pi_{\rm LO}$ is thus a local mass term, which we denote by $M^2$.
The exact propagator $D$ satisfies the equation
${\delta\Gamma/\delta D}=0$. In the present case, it reduces to a
local gap equation for $M^2$:
\bqa
M^2&=&m_4^2+\lambda
T\sum_{\omega_n=2\pi nT}\int_p{d^dq\over(2\pi)^d}
{1\over\omega_n^2+q^2+M^2}\;.
\label{gappi}
\eqa
The fields $\phi_0$ and $\rho_0$ satisfy 
the stationarity conditions 
${\delta \Gamma/\delta\phi_0}=0$
and ${\delta \Gamma/\delta\rho_0}=0$, i.e.
\bqa
\label{ss1}
0&=&m^2\phi_0-H+{\lambda\over2N}\phi_0\left(\phi_0^2+\rho_0^2\right)
+\lambda\phi_0T\sum_{\omega_n}\int_p{d^dq\over(2\pi)^d}
{1\over\omega_n^2+q^2+M^2}\;,\\
0&=&\left(m^2-\mu_I^2\right)\rho_0+{\lambda\over2N}\rho_0\left(\phi_0^2+\rho_0^2\right)
+\lambda\rho_0T\sum_{\omega_n}\int_p{d^dq\over(2\pi)^d}
{1\over\omega_n^2+q^2+M^2}\;.
\label{ss2}
\eqa
The gap equations require renormalization and details can be found in 
e.g.~\cite{jens}. The diagrammatic interpretation of the gap equations is
that they sum up all daisy and superdaisy diagrams.
The value of $M$ is found by
solving the renormalized versions of the 
equations~(\ref{gappi}),~(\ref{ss1}), and~(\ref{ss2}).
The quasiparticle masses are then found by solving the equation
$\det D(\omega,p=0)=0$, where the full propagator is obtained from 
the tree-level propagator Eq.~(\ref{d0}) by the replacement 
$m_i^2\rightarrow M_i^2=m_i^2+\Pi_{\rm LO}$. 

\section{Results}
\begin{figure}[htb]
\includegraphics[width=50mm]{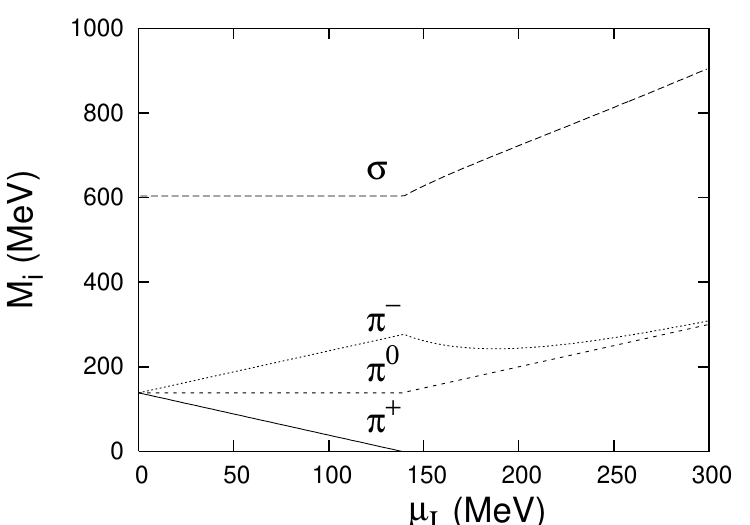}
\hspace{2cm}
\includegraphics[width=55mm]{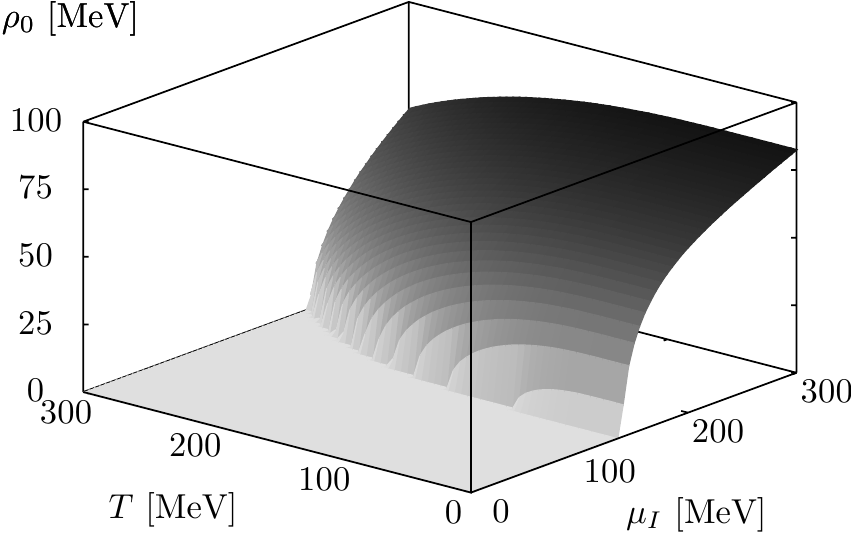}
\caption{Left: quasiparticle masses at $T=0$ at the physical point as a 
function
of $\mu_I$. Right: pion condensate at the physical point
as a function of $T$ and $\mu_I$.}
\label{masses}
\end{figure}

In Fig.~\ref{masses}, we show the quasiparticle masses as a function of
isospin chemical potential for $T=0$ and at the physical point.
The three pions are degenerate in the vacuum ($\mu_I=0$) and two of them
become degenerate in the limit $\mu_I\rightarrow\infty$. Also notice that 
$\pi^+$ becomes massless at the the critical value of 
$\mu_I=\mu_I^c=m_{\pi}=139$ MeV, which signals the onset of Bose condensation.
This is a Goldstone bosons arising from the
breaking of the $O(2)$-symmetry. 
In Fig.~\ref{masses}, we also show the pion condensate as a function of $T$
and $\mu_I$. Notice the onset of Bose condensation for $\mu_I=m_{\pi}$
at $T=0$. The phase transition is of second order in accordance with 
universality arguments.

\section{Summary and outlook}
\label{}
In this talk I have presented some results for Bose-Einstein condenstion
of charged pions in QCD at finite isospin chemical potential.
Using the 2PI $1/N$-expansion, I have calculated the quasiparticle masses
as well as the chiral condensate
and the pion condensate as a function of $T$ and $\mu_I$.
Our large-$N$ approximation is of mean-field type and
one would like to go beyond mean-field theory by including the next-to-leading
order in the $1/N$-expansion.
The appearance of massless excitations is in accordance with Goldstone's
theorem, which is respected by the 2PI 
$1/N$-expansion.

If Bose-condensation takes place in a star, the system must be
globally electric as well color neutral, otherwise one has to pay an 
enormous energy penalty. In the context of kaon condensation, one would
include electric charge neutrality by solving the equation
${\partial \cal \Gamma/\partial\mu_Q}=0$,
where $\mu_Q$ is the electric chemical potential.
The problem of kaon condensation in the CFL phase 
was investigated in detail 
using an $O(2)\times O(2)$ effective scalar
in the Hartree approximation
in the paper by Alford, Braby, and Schmitt~\cite{alfordus}, 
but without adressing
renormalization issues.
Work in this direction is in progress~\cite{lars}.
See also Ref.~\cite{fejos} for renormalization of
the 2PI Hartree approximation.


\section*{Acknowledgments:}
Part of the work presented in this talk was done in collaboration with
Tomas Brauner. 
The author would like to thank the organizers of SEWM2008 for 
a stimulating meeting.


\begin{thebibliography}{00}





\bibitem{alfordus}
M. Buballa, Phys. Lett. {\bf B609}, 57 (2005);
M. G. Alford, M. Braby, and A. Schmitt, J. Phys.
{\bf G}: nucl. Part. Phys. {\bf 35}, 025002 (2008);
M. G. Alford, A. Schmitt, K. Rajagopal, and T. Schafer, arXiv:0709.4635,
to appear in Rev. Mod. Phys.
\bibitem{andreas} A. Schmitt, these Proceedings.


\bibitem{lattqcd}
J. B. Kogut and D. K. Sinclair, 
Phys. Rev. D {\bf 64} 014508 (2002); ibid D {\bf 66} 34505 (2002);
S. Gupta, hep-lat/0202005;
P. de Forcrand, M. A. Stephanov, U. Wenger,
PoS LAT2007, (237) 2007. 
\bibitem{cpt}
K. Splittorff, D. T. Son and M. Stephanov, Phys. Rev. D {\bf 64} 016003 (2001).
J. B. Kogut and D. Toublan, Phys. Rev. D {\bf 64} 034007 (2001);
M. Loewe and Villavicencio, Phys. Rev. D {\bf 67}, 074034 (2003); ibid D 
{\bf 70}, 074005 (2004); ibid D {\bf 71}, 094001 (2005). 
\bibitem{njlmod}A. Barducci, R. Casalbuoni, G. Pettini, and L. Ravagli, 
Phys. Rev. D {\bf 69}, 096004 (2004):
D. Ebert and K. G. Klimenko, 
Eur. Phys. J. {\bf C46}, 771 (2006); 
J. O. Andersen and L. Kyllingstad, hep-ph/0701033;
D. Zablocki, D. Blaschke, and R. Anglani,
arXiv:0805.2687.
H. Abuki, R. Anglani, R. Gatto, M. Pellicoro, and M. Ruggieri,
arXiv:0809.2658.
\bibitem{china}L. He, M. Jin, and P. Zhuang, Phys.Rev. D {\bf 71}, 116001 
(2005).
\bibitem{pnjl}
S. Mukherjee, M. G. Mustafa, and R. Ray, Phys. Rev. D {\bf 75}, 094015 (2007); 
C. H. Abuki, M. Ciminale, R. Gatto, G. Nardulli, and M. Ruggieri,
Phys. Rev.  D {\bf 77}, 014040 (2008).
\bibitem{jens}J. O. Andersen, Phys. Rev. D, {\bf 75}, 065011 (2007).
\bibitem{jenstom}J. O. Andersen and T. Brauner, 
Phys. Rev. D, {\bf 78}, 014030 (2008).


\bibitem{lars} J. O. Andersen and L. Leganger, in preparation.
\bibitem{fejos}
J. Berges, Sz. Borsanyi, Urko Reinosa, and J. Serreau,
Annals  Phys. {\bf 320}, 344 (2005);
G. Fejos, A. Patkos, and Zs. Szep, Nucl. Phys. 
{\bf A803}, 115 (2008); A. Patkos and Zs. Szep, 
arXiv:0809.3551 (these Proceedings);
A. Jakovac, arXiv:0808.1800 and these Proceedings.

\end{thebibliography}
\end{document}